# Impact of Free Carriers on Modulational Instability in Silicon-on-insulator Nanowaveguides

**Deepa Chaturvedi\* and Ajit Kumar**
*Department of Physics, Indian Institute of Technology Delhi, New Delhi-110016, India*
*\*Corresponding author: waytodeepa@gmail.com*

**Abstract**: We have numerically studied the effect of free-carrier-induced loss and dispersion on the modulational instability (MI) gain at low input powers in silicon-on-insulator (SOI) nanowaveguides with normal and anomalous second-order dispersion. We have shown that the free carriers affect the gain spectra even at low input powers. First time we have reported the gain in normal SOI nanowaveguides even in the absence of higher order dispersion parameters, which is due to the interaction of free-carrier-induced dispersion and nonlinearity. The MI gain in an anomalous SOI nanowaveguide vanishes even at a few milliwatt range of input power due to this interaction. We have shown that the gain could be achieved in an anomalous nanowaveguides by reducing the free carrier lifetime.

## 1. Introduction

Modulational instability (MI) is one of the basic phenomena in nonlinear physics [1-3] which has attracted much attention due to its applications in high speed optical communication system like pulse train generation with ultrahigh repetition rate [4-6], supercontinuum generation [7,8] and wavelength conversion [9]. MI manifests itself in the exponential growth of any perturbation in the amplitude or/and phase of a continuous background field. It is due to the dynamical interplay between the nonlinear and dispersive or nonlinear and diffractive properties of the medium. The former is called the temporal modulation instability and latter one is spatial modulation instability. MI has been extensively studied in nonlinear optics, since it was first observed by K. Tai et al. in 1986 in single mode optical fiber [10], like in layered nonlinear cubical media [11], fiber amplifiers [12,13], two core optical fibers [14], tapered optical fiber [15], fiber bragg gratings [16] directional coupler [17], nanowires [18], metamaterials [19,20] etc.

In recent years, silicon-on-insulator (SOI) waveguide has become an important area of research in Silicon Photonics, as it can be integrated with complementary metal-oxide semiconductor (CMOS) technologies [21], and it shows strong mode confinement due to a large refractive index contrast of silica and silicon which leads to ultra-compact inexpensive optical devices [22, 23]. The strong nonlinearity of SOI nanowaveguide ensures that MI occurs easily and at a relatively shorter distance of propagation. It has been shown [24] that, in comparison with the optical fiber, an enhancement of two to three orders of magnitude in MI gain can be achieved in a SOI waveguide for an identical input peak power. This has resulted in a series of research studies on MI in SOI waveguides after Nicolae C. P. et al. did in 2006 [18]. W. Zhou studied the effect of linear loss and linear loss with two photon absorption [24, 25] on MI in SOI waveguides. Moreover, dispersion and nonlinearity tailoring in Silicon slot waveguide structure [26] and MI in SOI directional coupler [27] are also studied. Further, the compactness of SOI waveguide devices is studied in 1997 [28], and fabricated ultra-compact corner mirrors and T-Branches [29] and integrated waveguide turning mirrors [30] in SOI waveguides.

Silicon supports two-photon absorption (TPA) at telecommunication wavelengths due to an indirect band gap of 1.12 eV [31]. TPA generates free carriers, which further absorb light leading to enhanced absorption and also modify the refractive index of silicon. The former is known as free carrier absorption or FCA, while the latter is called free carrier dispersion or FCD. We thus see that generation of free carriers does considerably change the physical properties of a SOI waveguide and might impose some limitations on nonlinear silicon photonic devices. Therefore, it is important to study the role played by FCA and FCD in the context of modulational instability in SOI waveguides. However, in previous literatures, the free carriers (FC) effect is neglected completely at low input powers [18, 24, 25], which obviously impacts the gain spectra of MI.

The given work is dedicated to the study of the effect free-carriers (FC) on temporal MI at low powers in silicon-on-insulator (SOI) nanowaveguides with normal and anomalous second-order dispersion. Henceforth, we call a SOI waveguide as normal, if it has a positive group velocity dispersion (GVD) and we call it anomalous, if GVD is negative for it. Similar to optical fibers [10, 32-34], the MI gain spectra is supported in anomalous dispersion regime in SOI waveguides [24, 25]. It is possible in normal region with co-propagating one more wave [18]. For the first time we are reporting the existence of gain spectra in normal SOI waveguides due to free carriers, although the free carriers are limiting the gain in anomalous waveguides. We have shown that the gain in anomalous waveguide is achieved by reducing the free carrier lifetime. We also studied the role played by the free-carrier lifetime on the evolution of MI.

## 2. Theoretical Analysis

The dynamics of a pulse, propagating along the positive z direction in a SOI waveguide under the influence of group velocity dispersion, Kerr nonlinearity, two-photon absorption, FCA and FCD, is governed by the following set of coupled nonlinear differential equations [18, 35]. We have neglected loss to make our calculation simple.

$$\frac{\partial A}{\partial z} + i\frac{\beta_2}{2}\frac{\partial^2 A}{\partial T^2} = -\frac{ck\beta_1}{2n_0}\alpha_{fc}A + i\frac{\omega_0 k\beta_1}{n_0}n_{fc}A + i\frac{3\omega_0\beta_1^2}{\varepsilon_0 A_0}\Gamma|A|^2 A \qquad (1)$$

$$\frac{\partial N}{\partial T} = \frac{3\beta_1^2}{\varepsilon_0 \hbar A_0^2}\Gamma''|A|^4 - \frac{N(z,T)}{\tau} \qquad (2)$$

In the above equations, $A = A(z, T)$ is the slowly varying electric field amplitude, $z$ is the transmission distance, $T$ is the time in the frame moving with the group velocity $v_g$, $n_0$ is the refractive index of silicon, $\omega_0$ is the carrier frequency, $A_0$ is the cross-sectional area of silicon. The dispersion is contributed by first and second order dispersion parameters (group velocity dispersion, GVD) i.e. $\beta_1$ and $\beta_2$ respectively. The nonlinearity is attributed by self-phase modulation and two photon absorption, which is explained by the overlap integral of waveguide modes and third order linear susceptibility tensor, and is denoted by $\Gamma$. The fractional power stored in the core is defined by the parameter $k$. The parameters $\Gamma$ and $k$ are given by

$$\Gamma = A_0 \int_{A0} \hat{e}^* \cdot \chi^{(3)} : \hat{e}\hat{e}^*\hat{e} \, dA \Big/ \left( \int_{A\infty} n^2(r_t)|\hat{e}|^2 \, dA \right)^2 \qquad (3)$$

$$k = n_0^2 \int_{A0} |\hat{e}|^2 \, dA \Big/ \int_{A\infty} n^2(r_t)|\hat{e}|^2 \, dA \qquad (4)$$

Here $\chi^{(3)}$ is the third order susceptibility which depends on the symmetry of the waveguide and hence $\hat{\chi}^{(3)}_{1111} = (5.21 + i0.26) \times 10^{-19} m^2/V^2$, $\hat{\chi}^{(3)}_{1122} = (2.5 + i0.18) \times 10^{-19} m^2/V^2$ and $\hat{e}$ is the waveguide mode. The complex nature of $\chi^{(3)}$ make $\Gamma$ complex, where the real part denotes self-phase modulation ( SPM ) and the imaginary part denotes TPA coefficient. Free-carrier dynamics is explained by eq. (2), where $\tau$ is the free carrier lifetime, $n_{fc}$ is the change in the effective index due to FC-induced dispersion, $\alpha_{fc}$ is the FC-induced loss coefficient, $N_e$ and $N_h$ are the electron and hole density created due to TPA phenomena. Since the density of electrons and holes created due to TPA are same which can be denoted as $N$ and hence $\alpha_{fc}$ and $n_{fc}$ are given by [23, 36, 37]

$$\frac{\alpha_{fc}}{cm} = \left[8.5 \times \frac{N_e}{cm^3} + 6.0 \times \frac{N_h}{cm^3}\right] \times 10^{-18} = 14.5 \times \frac{N}{cm^3} \times 10^{-18}, \qquad (5)$$

$$n_{fc} = -\left[8.8 \times 10^{-4} \times \frac{N_e}{cm^3} + 8.5 \times \left(\frac{N_h}{cm^3}\right)^{0.8}\right] \times 10^{-18} = -\left[8.8 \times 10^{-4} \times \frac{N}{cm^3} + 8.5 \times \left(\frac{N}{cm^3}\right)^{0.8}\right] \times 10^{-18}. \qquad (6)$$

## 3. Linear Stability Analysis

The coupled nonlinear Schrodinger equation can be written as

$$\frac{\partial A}{\partial z} + i\frac{\beta_2}{2}\frac{\partial^2 A}{\partial T^2} = -a_1\alpha_{fc}A + ia_2\Delta n_{fc}A + ia_3\{a_4 + ia_5\}|A|^2 A \tag{7}$$

$$\frac{\partial N}{\partial t} = a_6|A|^4 - \frac{N(z,t)}{\tau} \tag{8}$$

where,

$$a_1 = \frac{ck\beta_1}{2n_0} \quad a_2 = \frac{\omega_0 k\beta_1}{n_0} \quad a_3 = \frac{3\omega_0\beta_1^2}{\varepsilon_0 A_0} \quad a_4 = R(\Gamma) \quad a_5 = \text{Im}(\Gamma) \quad a_6 = \frac{3\beta_1^2}{\varepsilon_0 \hbar A_0^2}\Gamma'' \tag{9}$$

The steady state solutions of the above system of differential equations are

$$\phi_{ss} = \phi_0 \exp[-\sigma^0 z + i\rho^0 z]$$

$$N_0 = a_6 \tau P_0^2 \exp(-4\sigma^0 z), \tag{10}$$

where $\sigma^0 = a_1\alpha_{fc}^0 + a_3 a_5 P_0$, $\rho^0 = a_2 n_{fc}^0 + a_3 a_4 P_0$, with $n_{fc}^0 = -\left[8.8\times 10^{-4} + (8.5\times 0.8)/\left(\frac{N_0}{m^3}\times 10^{-6}\right)^{0.2}\right]\times \frac{N_0}{m^3}\times 10^{-24}$,

$\frac{\alpha_{fc}^0}{m} = 14.5\times \frac{N_0}{m^3}\times 10^{-22}$, and $P_0 = |\phi_0|^2$ is the input peak power. To look at the linear stability of the steady state solutions, we add a small perturbation $f(z,T)$ to its amplitude and follow the evolution of $f(z,T)$ in accordance with the above set of dynamical equations. Hence, the perturbed solutions can be written as

$$\phi_{ss} = (\phi_0 + f)\exp[-\sigma^0 z + i\rho^0 z] \tag{11}$$
$$N = N_0 + \Delta N, \tag{12}$$

where $\Delta N$ is the corresponding perturbation to the free carrier density, which is calculated below. If we substitute (11) into (1) and keep only the terms linear in the perturbation, we obtain the following equation for the perturbing field $f(z,T)$:

$$\frac{\partial f}{\partial z} + i\frac{\beta_2}{2}\frac{\partial^2 f}{\partial T^2} = \left(-a_1\delta\alpha_{fc} + ia_2\delta n_{fc}\right)\sqrt{P_0} - ia_3(a_4 + ia_5)\left[P_0 f\left(1 - 2e^{-2\sigma^0 z}\right) - P_0 f^* e^{-2\sigma^0 z}\right]. \tag{13}$$

Here, $\delta n_{fc} = -\left[8.8\times 10^{-4} + (8.5\times 0.8)/\left(\frac{N_0}{m^3}\times 10^{-6}\right)^{0.2}\right]\times \frac{\Delta N}{m^3}\times 10^{-24}$, and $\frac{\delta\alpha_{fc}}{m} = 14.5\times \frac{\Delta N}{m^3}\times 10^{-22}$ are the perturbations in the free-carrier-induced dispersion and loss, respectively. To calculate these quantities, we have to first get $\Delta N$. It is given by

$$\Delta N = 2a_6 \tau P_0 \sqrt{P_0}(f + f^*)e^{-4\sigma^0 z} \tag{14}$$

As a consequence, we obtain

$$\delta\alpha_{fc} = 14.5\times 10^{-22} \times 2a_6\tau P_0\sqrt{P_0}\, e^{-4\sigma^0 z}(f + f^*) = 2c_1\frac{N_0}{\sqrt{P_0}}(f + f^*); \tag{15}$$

$$\delta n_{fc} = -\left[8.8\times 10^{-4} + \frac{8.5\times 0.8}{(N_0\times 10^{-6})^{0.2}}\right]\times 10^{-24} \times 2a_6\tau P_0\sqrt{P_0}\, e^{-4\sigma^0 z}(f + f^*) = -2c_2\frac{N_0}{\sqrt{P_0}}(f + f^*), \tag{16}$$

with $c_1 = 14.5 \times 10^{-22}$ and $c_2 = \left\{ 8.8 \times 10^{-4} + \dfrac{8.5 \times 0.8}{\left(N_0 \times 10^{-6}\right)^{0.2}} \right\} \times 10^{-24}$.

From (13)-(16), we arrive at

$$\frac{\partial f}{\partial z} + i\frac{\beta_2}{2}\frac{\partial^2 f}{\partial T^2} = (-a_1 c_1 - i a_2 c_2)(f + f^*)2N_0 - i a_3(a_4 + i a_5)\left[ P_0 f\left(1 - 2e^{-2\sigma^0 z}\right) - P_0 f^* e^{-2\sigma^0 z} \right] \quad (17)$$

Hence, we get

$$\frac{\partial f}{\partial z} + i\frac{\beta_2}{2}\frac{\partial^2 f}{\partial T^2} = (y_1 + iy_2)f + (y_3 + iy_4)f^*, \quad (18)$$

where

$$\begin{aligned}
y_1 &= -2a_1 c_1 N_0 + a_3 a_5 P_0 \left(1 - 2e^{-2\sigma^0 z}\right) \\
y_2 &= -2a_2 c_2 N_0 - a_3 a_4 P_0 \left(1 - 2e^{-2\sigma^0 z}\right) \\
y_3 &= -2a_1 c_1 N_0 - a_3 a_5 P_0 e^{-2\sigma^0 z} \\
y_4 &= -2a_2 c_2 N_0 + a_3 a_4 P_0 e^{-2\sigma^0 z}.
\end{aligned} \quad (19)$$

Assuming, the weak perturbation have the form $f(z,T) = u(z,T) + iv(z,T)$ and separately taking real and imaginary parts, we obtain a sets of coupled equations

$$\begin{aligned}
\frac{\partial u}{\partial z} - \frac{\beta_2}{2}\frac{\partial^2 v}{\partial T^2} &= (y_1 + y_3)u - (y_2 - y_4)v \\
\frac{\partial v}{\partial z} + \frac{\beta_2}{2}\frac{\partial^2 u}{\partial T^2} &= (y_2 + y_4)u + (y_1 - y_3)v.
\end{aligned} \quad (20)$$

For harmonic perturbation, we put $\begin{pmatrix} u \\ v \end{pmatrix} = \begin{pmatrix} u_0 \\ v_0 \end{pmatrix} \exp[i(kz - \Omega T)]$, where $k$ is the wavenumber and $\Omega$ is angular frequency of the perturbing sinusoidal field. If we substitute it into (20), we obtain

$$\begin{aligned}
iku_0 + \frac{\beta_2 \Omega^2}{2} v_0 - (y_1 + y_3)u_0 + (y_2 - y_4)v_0 &= 0 \\
ikv_0 - \frac{\beta_2 \Omega^2}{2} u_0 - (y_2 + y_4)u_0 - (y_1 - y_3)v_0 &= 0
\end{aligned} \quad (21)$$

In the matrix notation, we can write

$$A \begin{pmatrix} u_0 \\ v_0 \end{pmatrix} = 0, \quad (22)$$

where

$$A = \begin{pmatrix} ik - (y_1 + y_3) & \dfrac{\beta_2 \Omega^2}{2} + (y_2 - y_4) \\ -\dfrac{\beta_2 \Omega^2}{2} - (y_2 + y_4) & ik - (y_1 - y_3) \end{pmatrix} \quad (23)$$

For the nontrivial solutions, the determinant of the system must be zero. The gain coefficient for the modulational instability will be given by

$$G = 2|\text{Img}(k)| \quad (24)$$

$$G = -2y_1 + 2\sqrt{y_3^2 - y_2^2 + y_4^2 - \frac{\beta_2^2 \Omega^4}{4} - \text{sgn}(\beta_2)|\beta_2|y_2\Omega^2} \quad (25)$$

## 4. Results and Discussions

We have considered a standard ridge waveguide structure of SOI nanowaveguides of length 1 mm, as shown in the figure 1. Silicon is a guiding material of the waveguide, whose width and height are defined by parameters W and H, respectively. The light is propagated along the length of the waveguide (z-direction). We have analyzed two different dimensions of SOI nanowaveguides which are experiencing opposite GVD parameters, to study the effects of FC's on MI. The GVD parameters of waveguides with dimensions, $W \times H = 400\,nm \times 220\,nm$ and $800\,nm \times 220\,nm$, are $-9.536\ ps^2/m$ and $8.7769 \times 10^{-2}\ ps^2/m$, respectively. Hence, the first waveguide is anomalous whereas the second is normal at wavelength $\lambda_0 = 1.55\ \mu m$. We have divided our results in two sub categories, i.e. anomalous and normal SOI nanowaveguides. Also, we have studied the maximum gain in these two nanowaveguides which is grouped in the third subcategory.

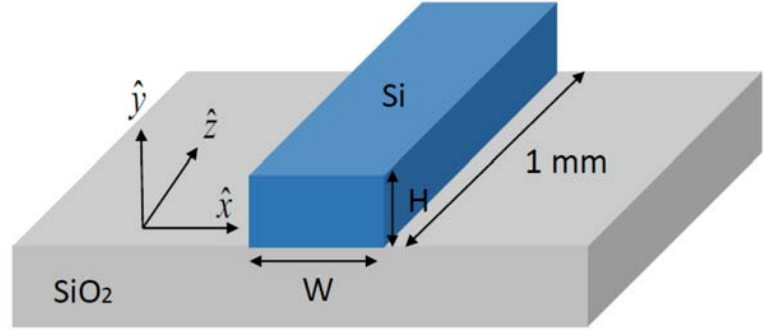

Figure 1: The structure diagram of SOI Nanowaveguide

4.1 Anomalous SOI Nanowaveguides

The parameters of an anomalous SOI nanowaveguide ($400\,nm \times 220\,nm$) are: $\beta_1 = 1.6167 \times 10^4\ ps/m$, $k = 0.9932$ and $\Gamma = 5.3626 \times 10^{-21} + \iota\, 2.6761 \times 10^{-22}\ m^2 V^{-2}$, which are calculated by perturbative method [38, 39]. The numerical values of other parameters used in the simulation are: $n_0 = 3.4764$ and the carrier lifetime $\tau = 0.5\,ns$ [18].

In the figure 2, we have shown the variation of MI gain with the distance of propagation and the frequency of modulation, in a 1 mm long anomalous SOI nanowaveguides for (a) 100 μW (b) 1 mW, (c) 10 mW and (d) 30 mW input peak powers, respectively in the absence of FC effects. It shows that the maximum gain of MI increases with the increase in the input peak power. We have shown the evolution of the gain spectra in the presence of FC effects for the same waveguide with same input powers in the figure 3. It clearly shows that the effect of FC's is visible in the milliwatt range of input peak power. When input power is increased to 1 mW, the maximum gain is reduced to approximately two-third of its value obtained without considering FC effects. Further by increasing the power, the gain spectra disappears. The change in the gain spectra is due to the combined effect of FC-induced dispersion and nonlinearity, which increases with the input peak power. The MI gain spectra for few 10's of milliwatt input peak power could also be achieved by controlling the carrier lifetime. Hence, we have also studied the influence of carrier lifetime on the MI gain in an anomalous nanowaveguide.

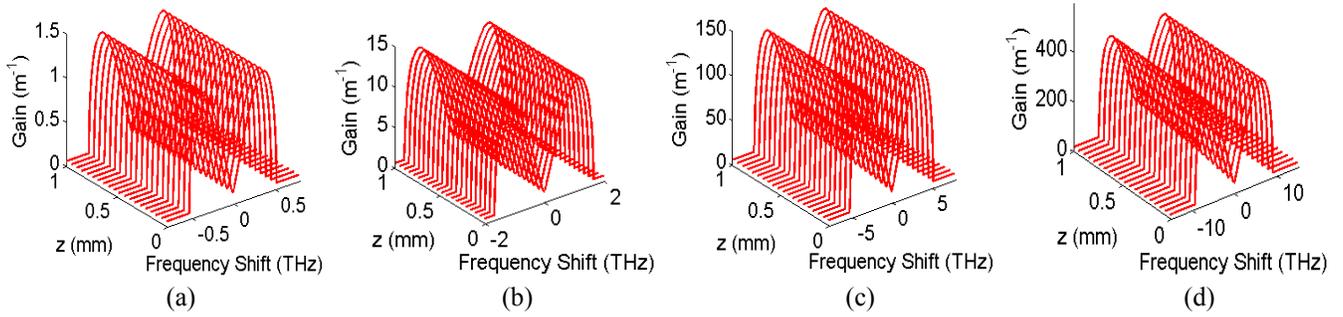

(a)                         (b)                     (c)                     (d)

Figure 2: Variation of the MI gain with the distance of propagation in a 1 mm long anomalous SOI nanowaveguide for (a) 100 μW, (b) 1 mW, (c) 10 mW and (d) 30 mW input peak powers, respectively when the FC-induced effects are ignored.

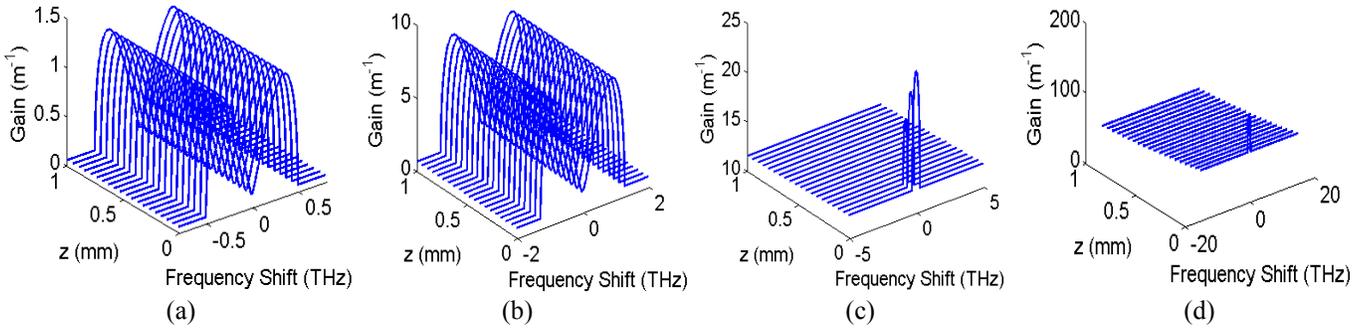

(a)                         (b)                     (c)                     (d)

Figure 3: Variation of the MI gain with the distance of propagation in a 1 mm long anomalous SOI nanowaveguide for (a) 100 μW, (b) 1 mW, (c) 10 mW and (d) 30 mW input peak powers, respectively when FC-induced effects are accounted.

### 4.1.1 The impact of carrier lifetime on MI gain

The MI gain disappears due to the interaction of FC-induced dispersion with nonlinearity as we have seen in the figures 3(c) and 3(d). Clearly, this detrimental effect can be reduced, if we could succeed in decreasing the carrier lifetime. It turns out that this can be achieved by silicon-ion implantation method [40]. Keeping this in mind, we have also studied the influence of free-carrier lifetime on the MI gain with an input peak power of 10 mW at 1 mm distance of propagation in the figure 4(a). It clearly shows that we can enhance the gain in an anomalous SOI nanowaveguide by reducing the carrier lifetime from the nanosecond range to the range of picosecond. Having determined this range of carrier lifetime, we have studied the effect of input peak power on the maximal gain in the figure 4(b) with a carrier lifetime of 1 ps and 10 ps. For a comparison, we have also shown the result for the case when FC-induced effects are absent. It clearly shows that the maximum gain in an anomalous waveguide could be achieved with a carrier lifetime of 1 ps, which is equivalent to the maximum gain in the absence of FC- induced effects.

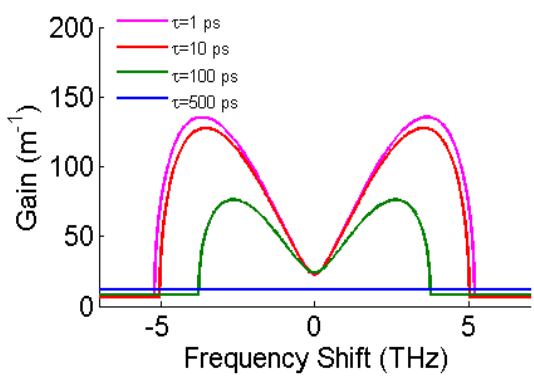 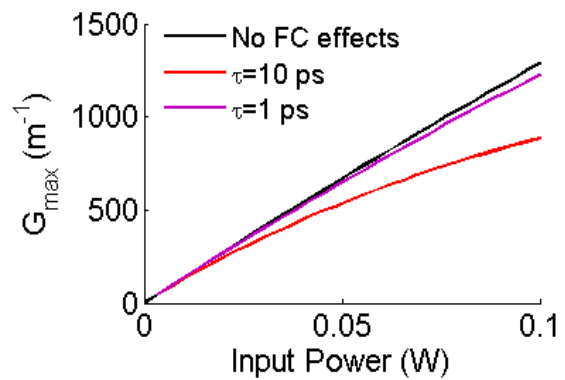

Figure 4(a): Variation of the MI gain with different values of carrier lifetime in anomalous SOI waveguide with an input peak power of 10 mW at 1 mm distance of propagation.

Figure 4(b): Variation of the maximum gain $G_{max}$ with an input peak power at 1 mm distance of propagation

## 4.2 Normal SOI Nanowaveguides

The parameters for simulation of normal SOI nanowaveguide ($800\,nm \times 220\,nm$) are: $\beta_1 = 1.3124 \times 10^4\,ps/m$, $k = 0.9924$ and $\Gamma = 6.4933 \times 10^{-21} + \iota 3.2404 \times 10^{-22}\,m^2 V^{-2}$, which are calculated by perturbative method.

In the figures 5, we have shown the variation of gain with the distance of propagation and the frequency of modulation, in a 1 mm long normal nanowaveguides for (a) 100 µW, (b) 1 mW, (c) 10 mW and (d) 30 mW input peak powers, respectively in the absence of free carrier effects. It shows that the MI does not exist in a normal SOI nanowaveguides with second order dispersion parameter. The figure 6 shows the MI gain spectra in a normal nanowaveguide for the respective power in the presence of FC effects. It shows that from microwatt range to 10 mW of input peak power, the gain spectra follows the same trend of existence of stability in the continuous-wave as that in the absence of FC effects. When the input peak power is 30 mW, the increased interaction of FC-induced dispersion with nonlinearity makes the cw-solution unstable and leads to the MI gain spectra as shown in the figure 6(d).

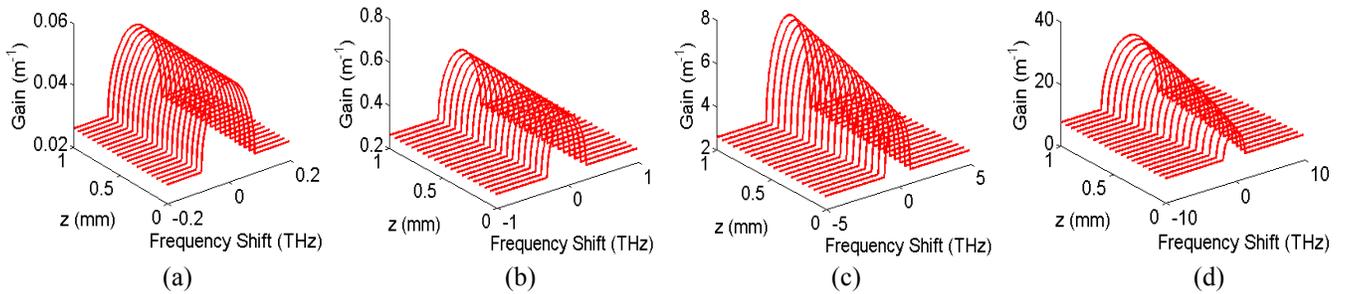

Figure 5: Variation of the MI gain with the distance of propagation in a 1 mm long normal SOI nanowaveguide for (a) 100 µW, (b) 1 mW, (c) 10 mW and (d) 30 mW input peak powers, respectively when the FC-induced effects are ignored.

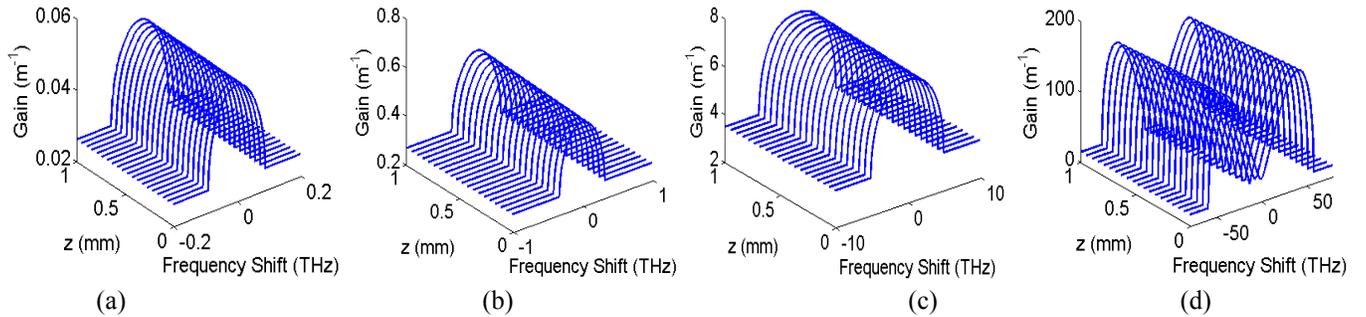

Figure 6: Variation of the MI gain with the distance of propagation in a 1 mm long normal SOI nanowaveguide for (a) 100 µW, (b) 1 mW, (c) 10 mW and (d) 30 mW input peak powers, respectively when FC-induced effects are accounted.

## 4.3 Maximum gain in anomalous and normal SOI nanowaveguides

To compare the maximum gain in an anomalous and normal SOI nanowaveguides, we have analysed two more SOI nanowaveguides, one anomalous ($500\,nm \times 220\,nm$) and other normal ($850\,nm \times 220\,nm$). In the figure 7, we have shown the maximum gain coefficient, $G_{max}$ in anomalous and normal SOI nanowaveguides. The free-carrier lifetime, in the case of anomalous nanowaveguides is 1 ps and in normal waveguide is 0.5 ns. The gain is maximum for a waveguide with smaller dimension in both the cases – anomalous or normal as in figures 7(a) and 7(b). It is clear that as the waveguide dimension increases, the nonlinearity decreases, and hence the resultant MI gain due to the interaction of dispersion with nonlinearity decreases.

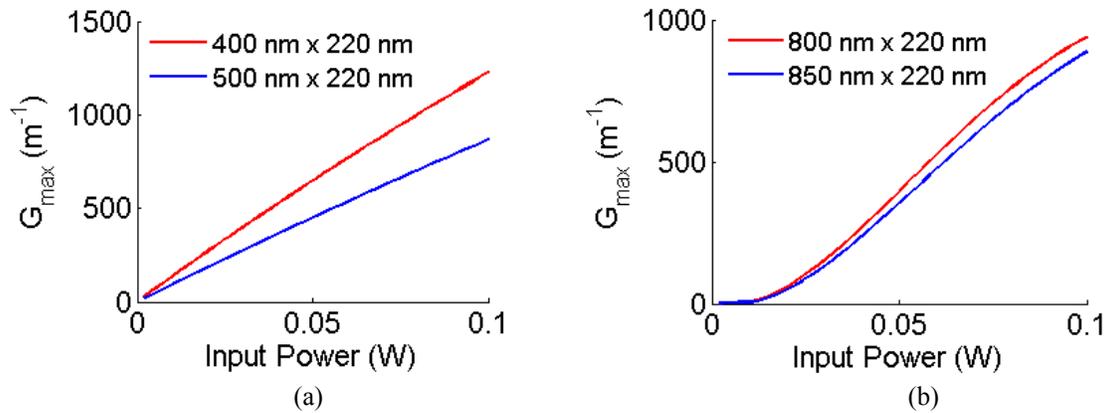

(a)                                         (b)

Figure 7: Variation of the maximum gain $G_{max}$ with an input peak power in (a) anomalous nanowaveguides and (b) normal waveguides.

## 5. Conclusions

We have numerically studied the phenomenon of modulational instability at relatively low input peak powers in an anomalous and normal SOI nanowaveguides. We have clearly demonstrated that, at the micro-Watt range of powers, the effect of FC on the MI gain spectrum is negligible, while it becomes effective at the milliwatt range of power due to increase in the interaction of FC-induced dispersion with nonlinearity. We have shown that the free carriers are responsible for the existence of the gain in the normal SOI nanowaveguides even in the absence of higher order dispersion parameters and disappearance of the gain in an anomalous nanowaveguides at few 10's of milliwatt range of input powers. We obtained that the MI gain in an anomalous nanowaveguide is achieved by reducing the free-carrier lifetime to 1 ps, which reduces the interaction of FC-induced dispersion with nonlinearity. The normal SOI nanowaveguides could also be used for terahertz pulse train generation, supercontinuum generation and wavelength conversion.

## 6. Acknowledgements

One of the author, Deepa Chaturvedi greatly acknowledges the award of Ph. D. fellowship from Council of Scientific and Industrial Research (CSIR India).